\documentclass[fleqn,usenatbib]{mnras}

\usepackage{newtxtext,newtxmath}

\usepackage[T1]{fontenc}

\DeclareRobustCommand{\VAN}[3]{#2}
\let\VANthebibliography\thebibliography
\def\thebibliography{\DeclareRobustCommand{\VAN}[3]{##3}\VANthebibliography}

\usepackage{graphicx}	
\usepackage{amsmath}	

\title[Optical and X-ray Follow-Up of GCU1]{Optical and X-ray Follow-Up to a Globular Cluster Ultraluminous X-ray Source in NGC 4472}

\author[Wasundara R. Athukoralalage]{Wasundara Ranhari Athukoralalage,$^{1}$\thanks{E-mail: athukora@msu.edu}
Kristen C. Dage,$^{2,3}$ 
Stephen E. Zepf,$^{1}$
\newauthor
Arash Bahramian,$^{4}$
Edward M. Cackett, $^{5}$
Arunav Kundu,$^{6}$
Thomas J. Maccarone$^{7}$
\\
$^{1}$Department  of  Physics  and  Astronomy,  Michigan  State  University,  East Lansing, MI 48824\\
$^{2}$ Department of Physics, McGill University, 3600 University Street, Montr\'eal, QC H3A 2T8, Canada\\
$^{3}$ McGill Space Institute, McGill University, 3550 University Street, Montr\'eal, QC H3A 2A7, Canada\\
$^{4}$International Centre for Radio Astronomy Research $-$ Curtin University, GPO Box U1987, Perth, WA 6845, Australia\\
$^{5}$Wayne State University, Department of Physics \& Astronomy, 666 W Hancock St, Detroit, MI 48201, USA \\
$^{6}$Eureka Scientific, Inc., 2452 Delmer Street, Suite 100 Oakland, CA 94602, USA\\
$^{7}$ Department of Physics \& Astronomy, Box 41051, Science Building, Texas Tech University, Lubbock, TX 79409-1051, USA \\
}

\date{Accepted XXX. Received YYY; in original form ZZZ}

\pubyear{2021}

\begin{document}
\label{firstpage}
\pagerange{\pageref{firstpage}--\pageref{lastpage}}
\maketitle

\begin{abstract}
NGC 4472 is home to five ultraluminous X-ray sources hosted by globular clusters. These sources have been suggested as good black hole candidates in extragalactic globular clusters$-$ a highly sought after population that may provide observational information regarding the progenitors of merging black hole binaries. In this body of work, we present X-ray and optical follow up to one of these sources, CXOUJ1229410+075744 (GCU1).
We find no evidence of [OIII] optical emission in GCU1 which indicates a lack of significant evidence for super-Eddington outflows, unlike what is seen in a handful of ULXs in extragalactic GCs. X-ray monitoring from 2019-2021 shows no detected X-ray emission above a few $\times$ $10^{38}$ erg/s. Comparisons of the multi-wavelength properties to disc-dominated, near Eddington Galactic black hole low mass X-ray binaries (GRS 1915+105 and XTEJ1817-330) suggests that GCU1 may show similar behaviour to GRS 1915+105 in terms of X-ray variability and similar relationships between $L_X$ and $kT$, with GCU1 showing maximum X-ray luminosities one order of magnitude higher.
\end{abstract}

\begin{keywords}
stars: black holes;
NGC 4472: globular clusters: individual;
X-rays: binaries
accretion, accretion discs
\end{keywords}



\section{Introduction}
Ultraluminous X-ray sources (ULXs) are bright X-ray binary sources that typically have luminosities exceeding $10^{39}$ erg/s (the Eddington Limit for a 10 $M_{\odot}$ black hole). While they were originally thought to be indicators of black holes, either $>10$ $M_{\odot}$ BHs accreting in standard physical regimes, or $<10$ $M_{\odot}$ BHs accreting in at super-Eddington luminosities \citep{gladstone2009}, the confirmation of a neutron star primary \citep{bachetti2014} in M82 X-2 has complicated the interpretation of these sources. 

A large number of ULXs are formed in the star forming regions of spiral galaxies \citep{Swartz2009}. However, a growing number of ULXs have been discovered in a very different stellar and dynamical environment, the globular clusters associated with early type galaxies \citep{Dage2019, Dage2020, Dage2021}. 
Because these ULXs are born in old and dynamic globular cluster systems, they are low mass X-ray binary (LMXB) ULXs, and undergo quite different formation mechanisms than their star-forming contemporaries. These key differences suggest that the best explanation for a GC ULX system is a black hole accretor. 

Given the rising interests in globular clusters as formation channels for binary black holes, such as those detected by LIGO \citep{Rodriguez2016}, and more theoretical work is showing that many black holes will be formed in clusters \citep{weatherford2020}, studying GC ULXs may be a fruitful way to build a sample of extragalactic globular cluster black hole candidates. 

CXOUJ1229410+075744 (hereafter ``GCU1") was first identified by \citet{Maccarone2011}. At the time it was one of only a few similar sources \citep{Maccarone2007, Irwin2010, Shih10}. Since then, up to 20 GC ULXs have been identified and studied in both X-ray and optical \citep{Roberts2012, Dage2019, Dage2020, Dage2021}. X-ray spectral studies of these objects probe the accretion processes fuelling the emission while the optical data constrain the properties of the host globular cluster.

Additionally, optical spectroscopy has also proven to be a useful tool in understanding these sources. \citet{Zepf2007, Zepf08} detected broad [OIII] emission above the globular cluster continuum in a NGC 4472 GC ULX, and \citet{Irwin2010} observed [OIII] and [NII] emission in a NGC 1399 GC ULX. Both of these sources showed similar X-ray behaviour, as they were both best-fit by a soft, thermal (kT$<$0.5 keV) spectrum which remained relatively constant as the sources varied in X-ray luminosity \citep{Dage2019b}. While only a handful of GC ULXs have measured optical spectra,  it is quite remarkable that the only two sources with optical emission showed very similar X-ray behaviour. This suggests that the presence or absence of optical emission, in combination with the X-ray behaviour, can potentially be used as a diagnostic for the accretion physics of the system.

GCU1 has been monitored in X-rays since the year 2000. It is well-fit by a single multi-color disc (MCD) \citep{mitsuda1984}  model, with no evidence for a hard component. The best-fit values for kT range from above 0.5 keV to just below 2.0 keV, and are strongly correlated to the X-ray luminosity. Its X-ray luminosity shows significant variability between observations, ranging from a few $\times 10^{37}$ erg/s to $\sim 3 \times 10^{39}$ erg/s \citep{Dage2019}.

X-ray variability is another key diagnostic in the study of extragalactic X-ray binaries. Previous studies of GCULXs have identified several different behaviours, with nine showing no variability within an observation or across many years of observations \citet{Dage2019, Dage2020, Dage2021}, eight that vary over the course of many observations, but not within an observation \citet{Dage2019, Dage2020}, and three that vary within an observation \citep{Maccarone2007, Shih10, Dage2020}. Variability is also seen in ULXs in star-forming galaxies, e.g. \citet{Earnshaw18} notes that transience in ULXs is generally uncommon but may be due to the propeller regime$-$where the accretion rate is stopped by a centrifugal barrier due to a neutron star's rotating magnetosphere \citep{Tsygankov2016}. Other ULXs (e.g. \citealt{2016ApJ...827L..13W}) may show 'super-orbital' periodicities thought to be caused by warped, precessing accretion discs \citep[e.g.][]{2001MNRAS.320..485O,2019ApJ...885..123H}. 

X-ray spectroscopy of ULXs can provide important diagnostics of
the physical mechanisms of their systems. Studies by \citet{Sutton2013} determine that ULXs are typically fit best by three main spectral types, and can undergo state changes (the broadened disc, hard ultraluminous, soft ultraluminous, as well as the ultrasoft ultraluminous state \citep{Urquhart2016}. Broadband spectroscopy of neutron star ULXs by \cite{Walton2018} found that the best-fit model included two thermal model components and one hard cutoff power-law component. \cite{Kobayashi19} explored model fits of ULXs likely to be black hole primaries and found that they were well described by a MCD model with a Comptonization component, and that the spectral properties could imply Bondi-Hoyle accretion. While GCU1 is located at a much further distance than many of these ULXs (d $\sim$ 16.8 Mpc), which translates to a much lower photon count rate and less detailed spectral features,  its lack of a hard emission component makes its properties different from those of vast majority of ULXs. While GCU1 might not be inherently best-fit by a MCD, the photons that do arrive suggest it is a disc-like source. Spectral studies of GCULXs in Virgo and Fornax (d$\sim$ 20 Mpc) find that they are best fit by a range of very soft MCDs to very hard power-law shapes \citep{Dage2019}.

Given that in the past, many studies of ULX behaviour have been compared to Galactic black holes \citep[e.g.][]{Miller2004}, we analyse new X-ray and optical observations of GCU1 and undertake the exercise of comparing GCU1's behaviour to near Eddington, disc dominated LMXB Galactic black holes GRS 1915+105 (hereafter GRS1915) and XTEJ1817-330 \citep{Rykoff2007, gierlinski08}. Section \ref{sec:analysis} describes the data and analysis methods, Section \ref{sec:results} discusses the results and comparison to GRS1915 and XTEJ1817-330, and the implications of this are presented in Section \ref{sec:discussion}.
\section{Data and Analysis}
\label{sec:analysis}
We analyse both X-ray and optical observations of GCU1. GCU1 has been observed in four \textit{Chandra} observations (PID 20620572, PI: Zepf) spanning 2019-2021 (see Table \ref{table:data}). It was observed on Gemini South/Gemini Multi-Object Spectrograph (GMOS) in 2011 (program GS-2012A-Q-57, PI: Zepf). 

\subsection{X-ray Analysis}
We reprocessed the X-ray images with \textsc{chandra\_repro} using CIAO version 4.12 \citep{Fruscione2006}, and filtered them to the 0.5-8.0 keV energy range using \textsc{dmcopy}. Although GCU1 has been detected in many past \textit{Chandra} observations, it was not detected in any of the new observations. We estimated non-detection upper limits by placing a 8 arcsecond region (as GCU1 is off-axis in the new observations by $\sim$ 3.3 arcmin) centred on GCU1's coordinates and measuring the counts in funtools \footnote{\url{https://github.com/ericmandel/funtools}}. Dividing by the exposure time and using $N_H$=1.6$\times 10^{20}$ cm$^{-2}$ and the best-fit power-law photon index ($\Gamma$=1.7) reported by \cite{Dage2019}'s fits of a power-law model to the observations, we used \textsc{pimms} \footnote{\url{https://cxc.harvard.edu/toolkit/pimms.jsp}} to convert the non-detection upper limit count to X-ray luminosities. The non-detection upper limit count rates and luminosities are presented in Table \ref{table:data}.

For the spectral fits to GCU1, we use \cite{Dage2019}'s measurements. The spectra with counts less than 100 were fit with \cite{cash} statistics, and the higher quality spectra were fit with $\chi^2$ statistics. GCU1 showed no statistical evidence for being better fit by either a single power-law model, or a combined power-law and disc model. For comparison, another GCULX source in the same galaxy and the same datasets, RZ2109, was statistically better fit by a disc plus power-law model \citep{dage2018}, which suggests that GCU1 is indeed best fit by a single disc model.

\subsection{Optical Data}
The colour and magnitude of GCU1 are $z$=20.8, $g-z$=1.59 \citep{Dage2019}. GCU1's optical spectrum was observed on Gemini via program GS-2012A-Q-57 with the 1.0 arcsec slit and B600 grating, resolution R=1688 \footnote{\url{https://www.gemini.edu/instrumentation/gmos/components\#Gratings}}. GCU1 was observed for 2 hours, 490 \AA \hspace{0.1cm}central wavelength and 1 hour with 510 \AA \hspace{0.1cm}central wavelength on February 25th, 2011, and for 2 hours with the 500 \AA\hspace{0.1 cm} central wavelength on February 29th, 2011. We reduced the spectrum using \textsc{IRAF} \citep{tody1986, tody1993}, and used \textsc{fxcor} \citep{tonry1979,alpaslan2009} to cross-correlate with a template M31 globular cluster spectrum over a wavelength range of 4600-5000 \AA \hspace{0.1 cm}(where the spectrum was relatively free of defects/cosmic rays) to measure radial velocity. GCU1's spectrum is shifted by 660$\pm$17 km/s relative to the rest frame. This is much greater than the radial velocities that are typical for foreground stars, and in the range of velocities used to classify objects as GCs for galaxies in the Virgo Cluster \citep{Strader2011}. 

\begin{figure}
\includegraphics[width=8cm]{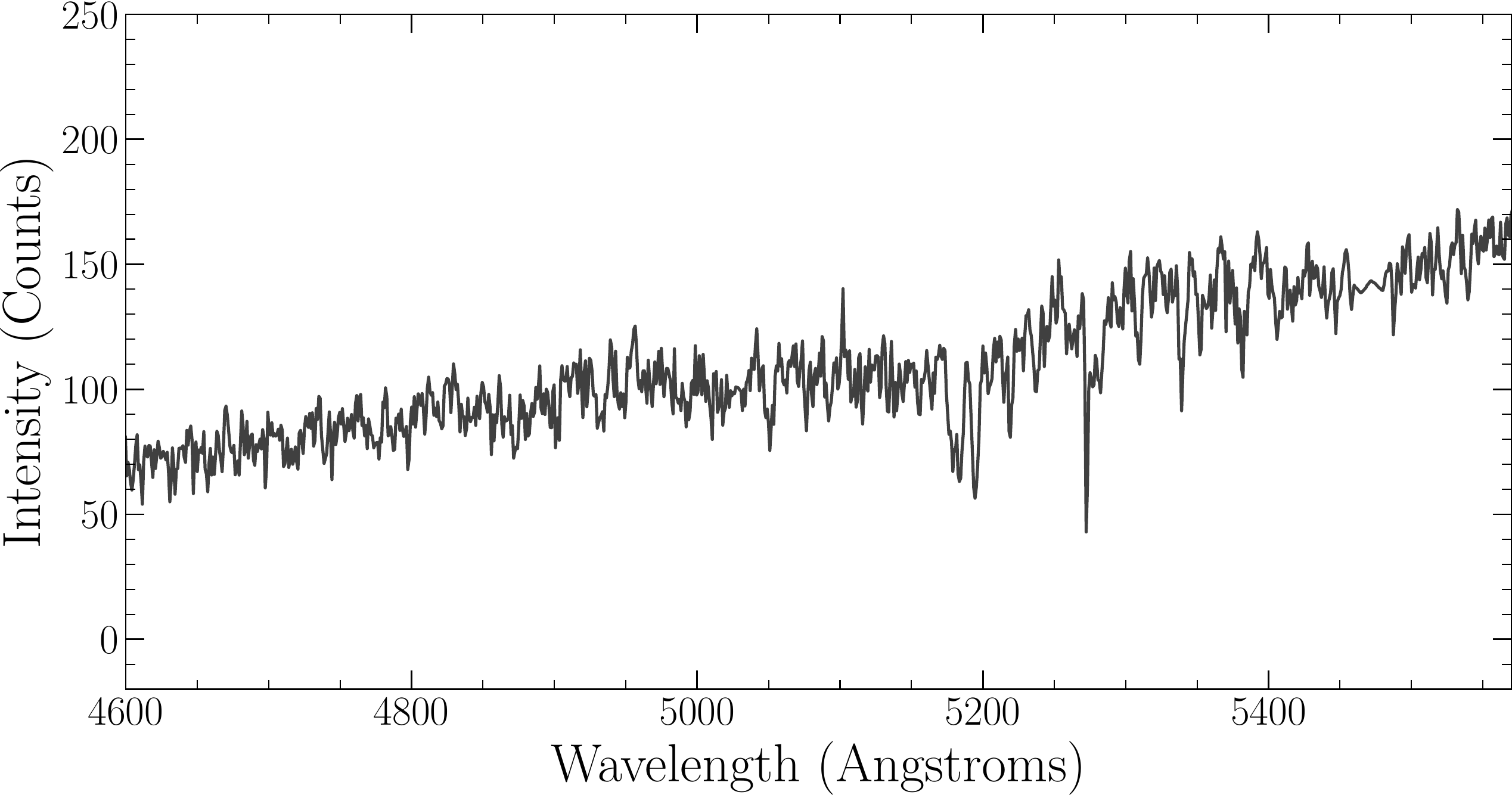}
\caption{Gemini South optical spectrum of GCU1 (CXOU1229410+075744), 500 \AA \hspace{0.1cm} central wavelength. The visible features in the spectrum are due to atmospheric induced sky-lines. }

\label{fig:spectrum}
\end{figure}

No significant optical emission above the cluster continuum was detected (see also Figure \ref{fig:spectrum}). Using the 500 \AA\hspace{0.1 cm} central wavelength combined spectrum, we search for evidence of [OIII] emission in the spectrum by computing the equivalent width (the relative brightness of the spectrum compared to the cluster continuum) in the regime of 5008-5028 \AA. We use a polynomial model of the GC continuum (see \citealt{Dage2019b} for more detail) spanning 4600-5150 \AA \hspace{0.1cm} (the regime where the spectrum is relatively free of chip defects/gaps). The continuum end-points were computed from 20 \AA \hspace{0.1cm}wide datapoints randomly subsampled from 4600:4630 \AA \hspace{0.1cm}and 5120:5150 \AA, as they did not have many features that deviated from the average value in those regimes.  We re-sampled this 50 times and find that the largest possible equivalent width is still less than $0.36$ \AA, with the average computed value typically less than 0.1 \AA. Both of these numbers are negligible when compared to detected [OIII] emission from other sources \citep[e.g.][]{Irwin2010, Dage2019b, Sun19}. For example, RZ2109's equivalent width ranges from 12.9 to 31.6 \AA \hspace{0.1cm}\citep{Dage2019b}, and \cite{Sun19} report equivalenth widths on the order of 4.0 \AA.  We could not search for [NII] emission as the optical spectrum did not extend above 6500 \AA.

\section{Results}
\label{sec:results}
We present new optical and X-ray measurements of the globular cluster ultraluminous X-ray source GCU1. We found that GCU1 showed significant X-ray variability in the past. In the 2019-2021 observations, GCU1 has not been detected. Analysis of GCU1's optical spectrum confirms the cluster nature of the counterpart, but no significant optical emission lines are detected beyond the cluster continuum (See Figure \ref{fig:spectrum}).

\subsection{Long-Term Lightcurve of GCU1 }

\begin{figure}
\includegraphics[width=9.5cm]{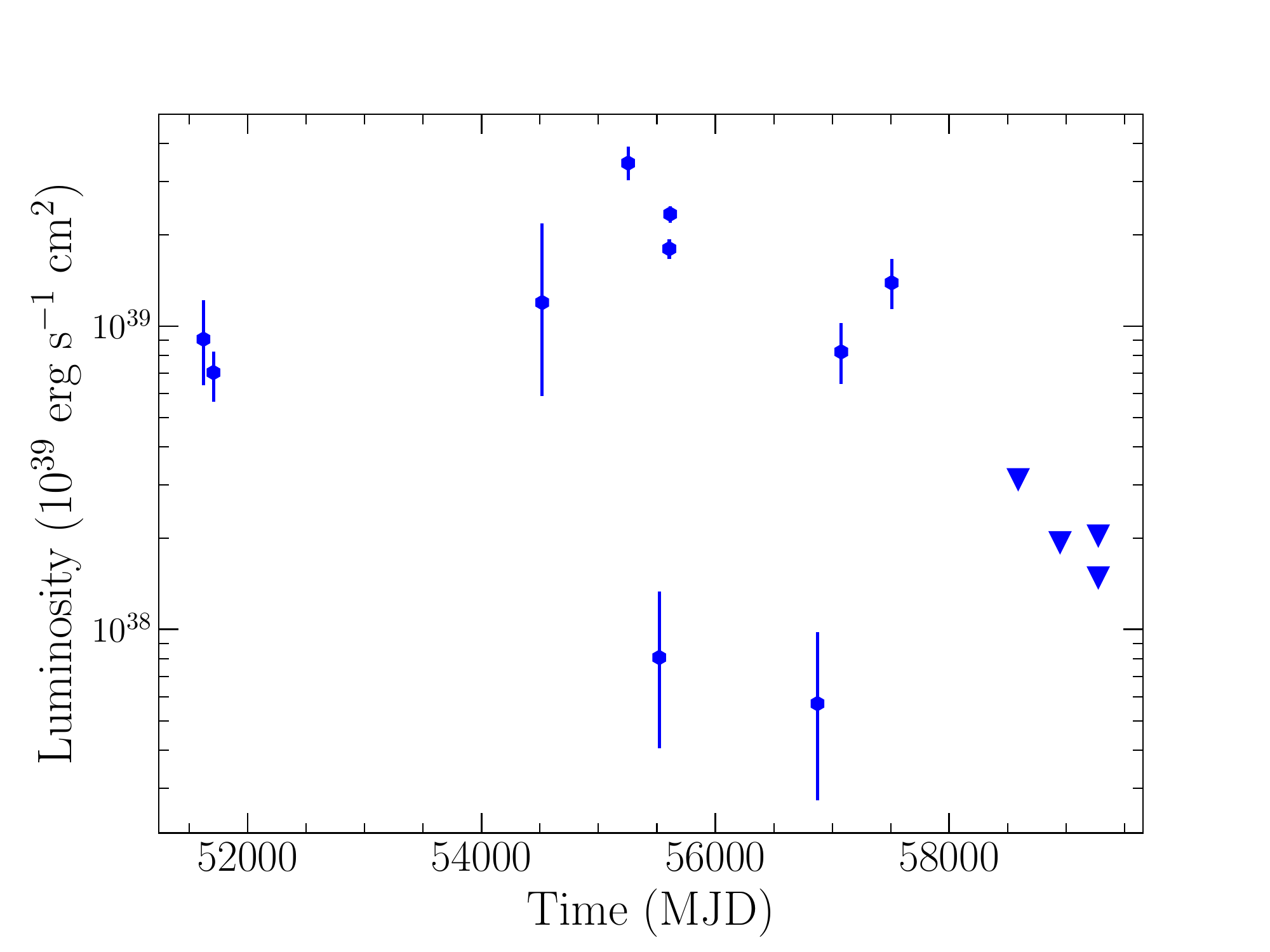}
\caption{ X-ray lightcurve of GCU1 from 2000-2021 (X-ray detections from \citet{Dage2019}, upper limits are presented in Table \ref{table:data}). While GCU1 has shown significant variability in the past observations, the source has not been detected above a few $\times 10^{38}$ erg/s in the most recent monitoring from 2019-2021.  }

\label{fig:lightcurve}
\end{figure}

The long-term X-ray behaviour of GCU1 can be seen in Figure \ref{fig:lightcurve}. Previous monitoring on a yearly timescale from 2014-2016 detected the source both above $10^{39}$ erg/s and below $10^{38}$ erg/s. The recent non-detections indicate that the source may have switched off in X-ray for the time being. GRS1915 is a Galactic BH (15 solar masses) and has been well-studied over the last 20 years, and has exhibited similar modes, as both exhibit high X-ray variability and switching-off \citep{Negoro2018,balakrishnan21}.

Analysis by \cite{2020ApJ...902..152N, Miller2020} suggests that the low state in GRS1915 is driven by an increase in the local absorption rather than a change in the intrinsic flux. As GCU1 is too far away to probe intrinsic absorption, it is not possible to determine whether the non-detections in GCU1 might be intrinsic or, like GRS1915, due to absorption. However, further comparisons in how the X-ray luminosity scales with inner disc temperature may be instructive to find clues regarding the nature of GCU1.

\subsection{A comparison of GCULX1 to GRS1915 and XTEJ1817-330}
\begin{figure}
\includegraphics[width=8cm]{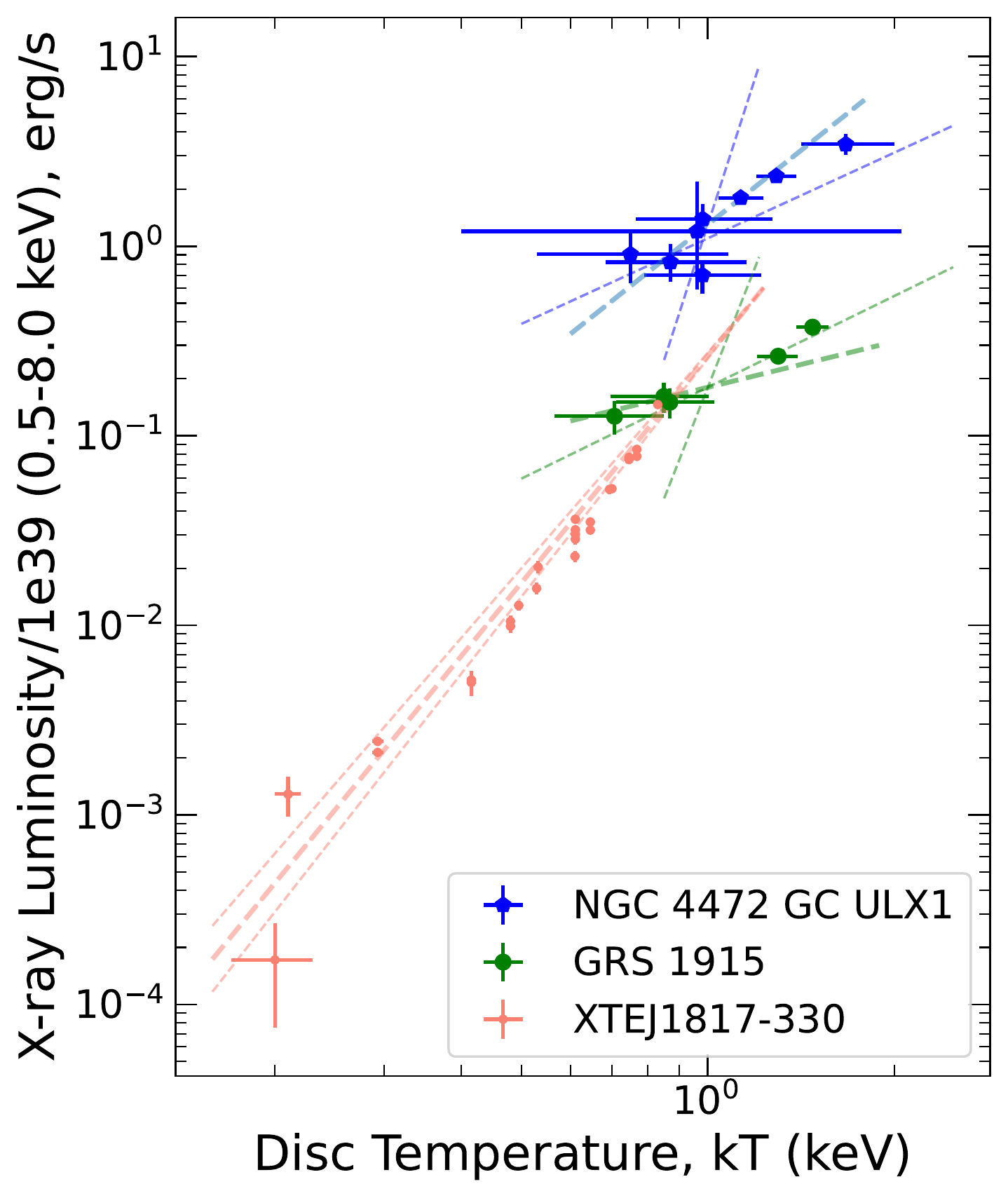}
\caption{X-ray luminosity versus $kT$ in log-scale for GCU1 as well as known Galactic black holes GRS1915. The data for GCU1 is measured in \citet{Dage2019}; the values for GRS1915 are digitized from \citet{Miller2004}, who use a distance of 11 kpc to calculate the luminosity. We apply the range of percent uncertainty in GCU1's measurements, sorted by luminosity, to GRS1915. The values for XTEJ1817-330 are taken from \citet{Rykoff2007}. We use \textsc{pimms} to convert the measurements from the 0.5-10 keV band to the 0.5-8.0 keV band for both GRS1915 and XTEJ1817-330. We overlay the data with the best fit slopes and uncertainties (dashed and dotted lines).}

\label{fig:lxg}
\end{figure}

We chose to compare GCU1 to two Galactic LMXBs based on the following criteria: GRS 1915 is an object in the same luminosity range as GCU1 with a well constrained distance and has been fit with the same \textsc{diskbb} model as GCU1 \citep{Miller2004}. XTEJ1817-330, in addition to also being fit with the \textsc{diskbb} model by \cite{Rykoff2007}, has been observed over a wider range of luminosities than GRS1915. 
We endeavour to perform a further comparison between GCU1 and GRS1915, although GCU1 is extragalactic (d=16.8 Mpc), by comparing how similar the X-ray luminosity/disc temperature trends are. Figure \ref{fig:lxg} shows the X-ray luminosity versus disc temperature for the two sources, which overall seem to show very similar trends. While GRS1915's true spectrum is complicated, we make this comparison using data digitized from \cite{Miller2004}, as they have been fitted to the same \textsc{diskbb} model used to fit the data of GCU1, and selected to probe a large range of temperatures. We use \textsc{pimms} to convert the measurements from the 0.5-10 keV band to the 0.5-8.0 keV band, and applied the percent uncertainties from GCU1, sorted by luminosity to GRS1915. We also converted the measurements from \cite{Rykoff2007} from the 0.3-10 keV band to 0.5-8.0 keV. We note that the nominal distance to XTEJ1817-330 is 5.5 kpc \citep{sala}, but that this measurement is highly uncertain. 

We use the package \textsc{linmix} \citep{Kelly07} to perform linear regression with a Bayesian approach while considering measurement uncertainties to fit the slopes of XTEJ1817-330, GRS1915, and GCU1 in log space to find the best-fit power-law index for the relationship between $L_X$ and $kT$. We found that the best fit index (and corresponding 1 $\sigma$ uncertainties) for GCU1 is 2.6 $^{+7.5}_{-1.1}$, and 1.6$^{+6.8}_{-0.8}$ for GRS1915. For XTEJ1817-330, we found a slope of 3.78$\pm 0.14$.
This analysis suggests that GCU1 and GRS1915 may show similar behaviour in $L_X$ and $kT$ space, although we caution that the observational data of GCU1 has large error bars and does not well constrain the slope.


Another method to estimate the lower limit of the black hole mass is by using the \textsc{diskbb} normalization. We take the measured normalizations from \cite{Dage2019}, and apply a correction to determine the disc radius from the apparent inner disc radius ( $ R_{in} = \kappa^2 \xi r_{in} \sqrt{\cos{\theta}} $; hardening ratio $\kappa$ =1.7 and correction factor $\xi=0.412$, see \citealt{shimura95, kubota98}), assuming a completely face-on disc ($\theta$=0). This results in an estimation of a mass lower limit of 10-20 $M_{\odot}$ $-$with corresponding inner disc radii estimates ranging from 500-1500 km$-$ although we caution that the measured normalizations are not well constrained, and a number of assumptions has gone into the estimation. 
    
\begin{figure*}
\centering
{{\includegraphics[width=7cm]{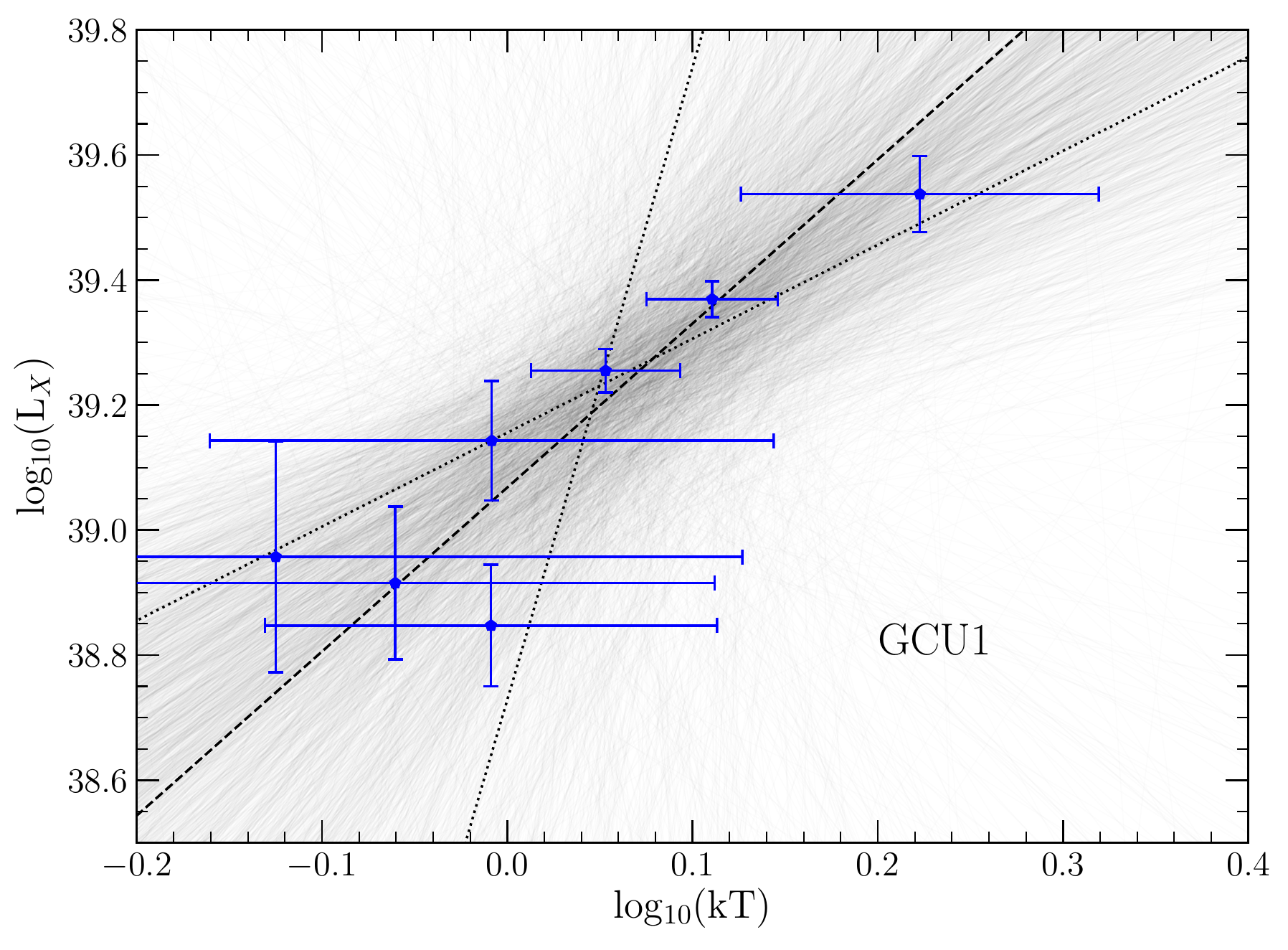} }}
\qquad
{{\includegraphics[width=7cm]{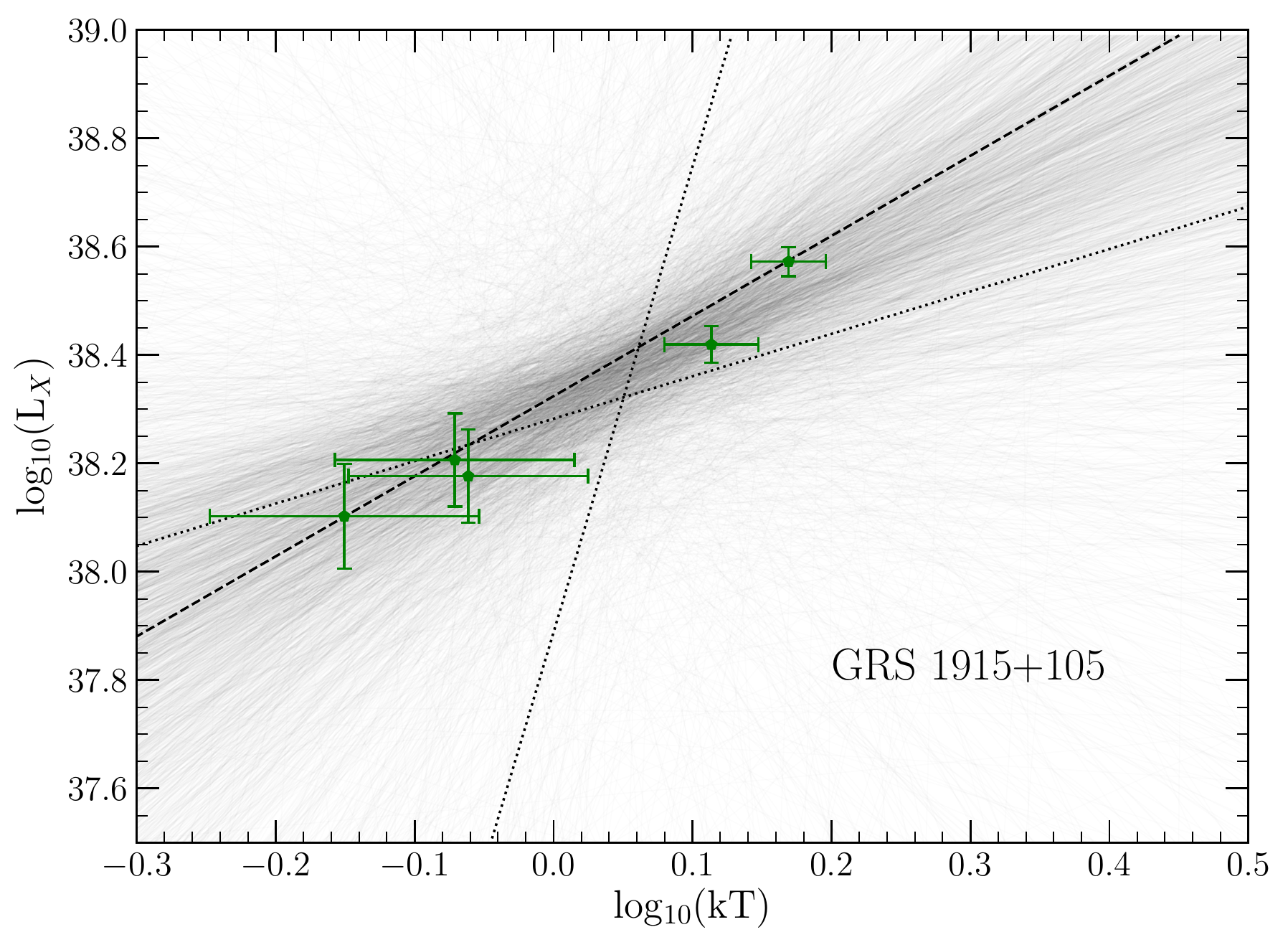} }}
 \caption{\textsc{linmix} slope fitting of GCU1 (left) and GRS1915 (right). The lines show the various fits from \textsc{linmix}, along with the best-fit and uncertainties (solid lines)}. 

 \label{fig:lxkt} 
\end{figure*}
\begin{figure*}
\centering
{{\includegraphics[width=7cm]{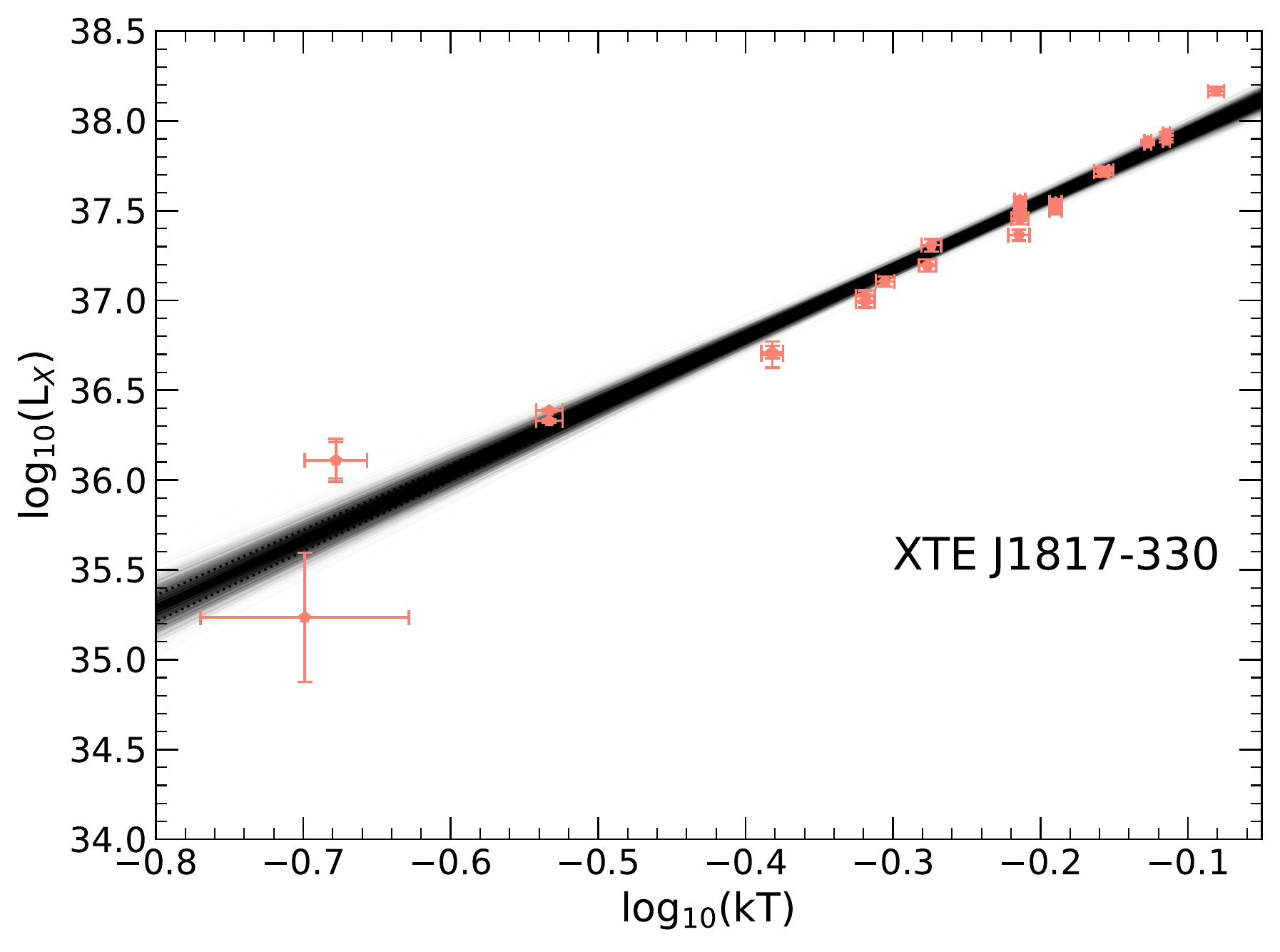} }}
\caption{\textsc{linmix} slope fitting of XTEJ1817-330. The lines show the various fits from \textsc{linmix}, along with the best-fit and uncertainties (solid lines)} 

 \label{fig:lxkt} 
\end{figure*}
 
\section{Discussion}
\label{sec:discussion}

We have analysed new X-ray and optical observations of the globular cluster ultra-luminous X-ray source GCU1. X-ray spectral analysis of ten GC ULXs by \cite{Dage2019} found that the sample best fit by a MCD fell into two different modes: a soft (kT $<$ 0.5 keV) disc where the X-ray luminosity spanned over an order of magnitude, but the best-fit temperature remained relatively unchanged, or best fit kT values $>$ 0.5 keV, with the luminosity increasing with the temperature. The ``soft'' sources in the first category both showed evidence of emission in the optical spectra, which implied that they were accreting in the super-Eddington regime. GCU1's X-ray spectrum fell into the latter category, with luminosity increasing with temperature, and GCU1's optical spectrum showed no evidence of emission, which implies that GCU1 is not under-going any major outflows, and may be in a near- or sub-Eddington accretion state.

The last twenty years worth of \textit{Chandra} monitoring observations of GCU1 has shown that the source is bright and highly variable. Our analysis of 2019-2021 data shows that GCU1 has ceased to be detected about luminosities greater than a few $\times 10^{38}$ erg/s. Both the inherent high variability of the source and ``turning off'' evokes modes of GRS1915.

While GCU1 is distant (d=16.8 Mpc), and the observable information is sparse, GCU1's overall trends show many similarities with GRS1915: a luminosity temperature relation and strong temporal variability. If these two sources are truly in similar accretion regimes, at sub-Eddington regimes, mass and luminosity are linearly correlated, which may imply that GCU1 may be somewhat more massive than GRS1915.

While it is difficult to make more conclusive results regarding the nature of GCU1, it is an important source to understand because it can provide clues of how extreme X-ray luminosities occur in old and crowded systems.

\begin{table}
\caption{\textit{Chandra} observations of GCU1 as well as count rate and 90\% confidence interval X-ray luminosity upper limits derived using \citet{Gehrels} statistics.}
\label{table:data}
\begin{tabular}{|l|l|l|l|l|}
\hline
\textbf{Obs ID} & \textbf{Date}& \textbf{Length} & \textbf{Count rate} & \textbf{$L_X$} \\
&&ks&  $10^{-3}$cts/s& $10^{38}$erg/s\\ \hline
21647  & 2019-04-17  & 29.68                &  $<$ 0.71                    &$<$ 3.13           \\ 
21648 & 2020-04-09 & 29.68                & $<$ 0.44                    &$<$ 1.94         \\ 
21649  & 2021-03-02  & 19.81                & $<$ 0.34                  & $<$ 1.49        \\ 
24981 & 	2021-03-08  & 11.48                &$<$ 0.46                    & $<$ 2.04          \\ \hline
\end{tabular}
\end{table}


\section*{Data Availability}
The \textit{Chandra} observations are available through \url{https://cda.harvard.edu/chaser/}, and the Gemini data is archived at \url{https://archive.gemini.edu/searchform}, program GS-2012A-Q-57. 

\section*{Acknowledgements}
The authors gratefully acknowledge the anonymous referee.
 KCD acknowledges funding from the Natural Sciences and Engineering Research Council of Canada (NSERC), fellowship funding from the McGill Space Institute, and from Fonds de Recherche du Qu\'ebec $-$ Nature et Technologies, Bourses de recherche postdoctorale B3X no. 319864. SEZ acknowledges support from grant GO9-20080X. This research has made use of data obtained from the Chandra Data Archive and the Chandra Source Catalog, and software provided by the Chandra X-ray Center (CXC) in the application package CIAO. Based on observations obtained at the international Gemini Observatory, a program of NSF’s NOIRLab, which is managed by the Association of Universities for Research in Astronomy (AURA) under a cooperative agreement with the National Science Foundation on behalf of the Gemini Observatory partnership: the National Science Foundation (United States), National Research Council (Canada), Agencia Nacional de Investigación y Desarrollo (Chile), Ministerio de Ciencia, Tecnología e Innovación (Argentina), Ministério da Ciência, Tecnologia, Inovações e Comunicações (Brazil), and Korea Astronomy and Space Science Institute (Republic of Korea). This work was performed in part at Aspen Center for Physics, which is supported by National Science Foundation grant PHY-1607611.
 


\newpage
\bibliographystyle{mnras}
\bibliography{ulxs4472} 





\bsp	
\label{lastpage}
\end{document}